# Strain Engineering of Correlated Charge-Ordered Phases in 1T-TaS$_2$

*Rafael Luque Merino[1,*], Felix Carrascoso[1], Eudomar Henríquez-Guerra[2], M. Reyes Calvo[2,3], Riccardo Frisenda[4] and Andres Castellanos-Gomez[1,*]*

[1]*2D Foundry Research Group. Instituto de Ciencia de Materiales de Madrid (ICMM-CSIC), Madrid, E28049, Spain*

[2]*BCMaterials, Basque Center for Materials, Applications and Nanostructures, 48940 Leioa, Spain;*

[3]*IKERBASQUE, Basque Foundation for Science, 48009 Bilbao, Spain*

[4]*Dipartimento di Fisica, Università di Roma "La Sapienza", I-00185 Roma, Italy*

*\*Corresponding Author*

rafael.luque@csic.es

andres.castellanos@csic.es

Strain engineering is a powerful strategy for controlling the structural and electronic properties of two-dimensional materials, particularly in systems hosting charge density wave (CDW) order. In this work, we apply uniaxial tensile and compressive strain to thin 1T-TaS$_2$ flakes using a flexible, device-compatible platform, and systematically investigate the strain-dependent behavior of the nearly commensurate (NC) to incommensurate (IC) CDW phase transition. This transition is driven by Joule heating at room temperature. Electrical transport measurements reveal that both the switching threshold voltage and the resistance of the NC-CDW phase exhibit clear, reversible strain dependence. Furthermore, we identify a quadratic dependence between the strain-induced resistance change and the threshold voltage, confirming that piezoresistive modulation governs the strain-tunability of the phase transition. We demonstrate a room-temperature, electrically-readout strain and displacement sensor with threshold-like response in a programmable window. These results highlight the potential of 1T-TaS$_2$ for on-chip sensing of strain and displacement.

Charge density waves (CDW) are periodic modulations of the electronic charge density, coupled to lattice distortions[1]. CDWs generally emerge due to electron-electron interactions (via Fermi surface nesting)[1,2] and electron-phonon interactions (with an associated CDW phonon mode)[3,4]. Materials hosting CDW order are of great interest due to their rich phase diagrams, where the charge order is intimately linked to correlated, low-temperature phases like superconductivity[5,6] and Mott physics[7–9]. Two-dimensional (2D) materials hosting CDWs are particularly appealing, as charge order can persist above room temperature[10] (RT) and the reduced dimensionality boosts their susceptibility to external perturbations[11].

Strain engineering in 2D materials provides a versatile platform to modulate materia properties: from bandgap engineering leading to changes of electronic and optical properties[12]; to tipping the balance between ground states in correlated 2D materials[13–16]. Within correlated materials, strain constitutes a natural tuning knob for CDW materials, as the charge order itself is coupled to a lattice distortion and thus highly sensitive to mechanical deformation. Strain engineering of CDW materials has been shown to alter the ground state[7,17], enhance or suppress charge order[18–22] or modify the dynamics of metastable states[23–25]. However, most studies rely on bulky hardware incompatible with integrated devices[26–29]. Wrinkling via patterned substrates offers an on-chip alternative, yet it imposes only fixed, non-tunable strain[30,31], similar to unintentional strain resulting from growth processes[32]. These limitations highlight the opportunity to harness large, well-defined strain in device-compatible geometries to manipulate CDW order *in situ* and unlock novel electronic functionalities.

1T-TaS$_2$, a layered transition metal dichalcogenide, is an attractive platform for tailoring CDW order because it hosts a hierarchy of CDW phases: from incommensurate (IC) order at high temperatures, to nearly commensurate (NC), to fully commensurate (C) order at low temperatures. Extensive work demonstrates that the individual CDW phases, and the transitions linking them, can be actively manipulated through a multitude of external perturbations, such as electrostatic gating[33,34], electromagnetic fields[35–43], targeted chemical doping[44–46] and others[47,48]. Of particular relevance for applications, the NC-to-IC phase transition occurs around $T_C \approx 350$ K, opening the door to practical devices that exploit CDW phase transitions near RT[35,36,40,41,49,50].

Previous studies of strain in 1T-TaS$_2$, many of which focused on hydrostatic pressure[28,51], revealed dramatic effects such as a collapse of the Mott gap[52,53], control over metastable CDW phases[31,54], or the emergence of superconductivity[28,51]. A handful of experiments have also explored in-plane deformations, but they rely on complex and resource-intensive setups[27–29] or fixed-geometry approaches[30,31] that preclude dynamic tuning. Crucially, most reports target the low-temperature phases, leaving the NC and IC orders near 350 K largely unexamined under strain. Dynamical strain-tuning of this phase transition could enable room-temperature, device-compatible CDW functionalities.

In this work, we employ a simple yet versatile method to apply uniaxial strain continuously and reversibly, enabling dynamic control of the NC-to-IC CDW transition in 1T-TaS$_2$. This approach not only provides a practical route for exploring strain-dependent behavior but also lays the foundation for the development of strain-sensitive devices. To illustrate this potential, we demonstrate two room-temperature electrical detection of both tensile and compressive strain in two distinct operation modes. Remarkably, one can leverage the CDW phase transition as an intrinsic amplification mechanism to obtain exceptionally large strain sensitivity.

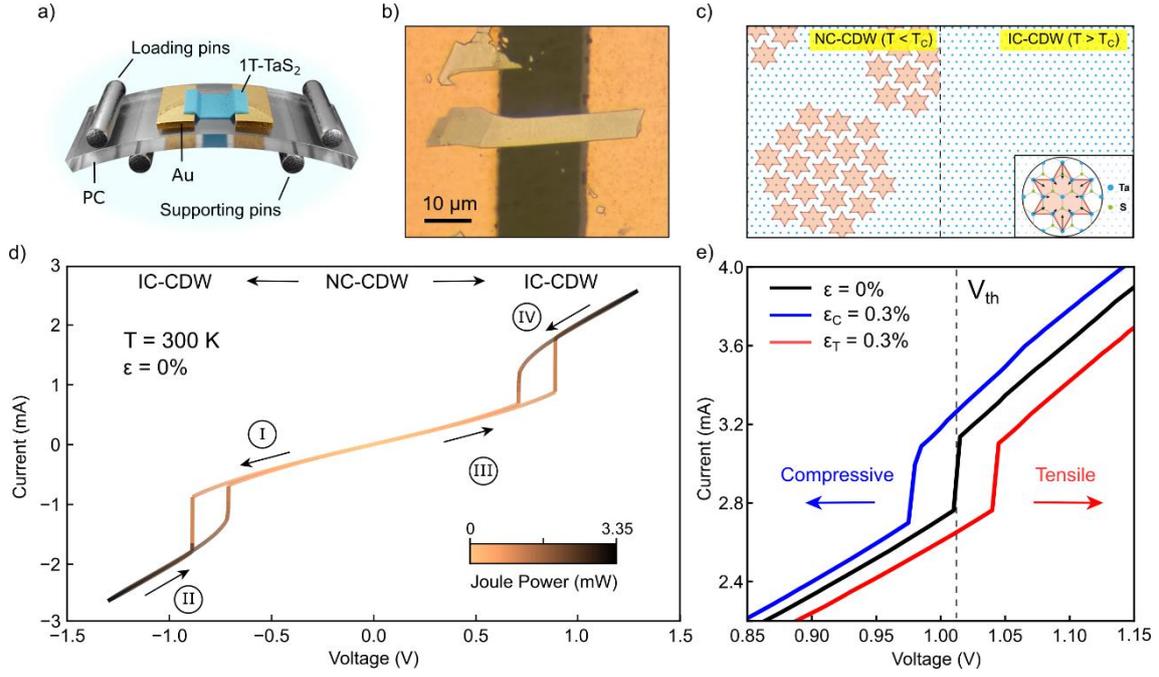

**Figure 1.** a) Experimental concept: a two-terminal 1T-TaS$_2$ device fabricated on polycarbonate (PC) is uniaxially strained using a four-point bending setup. b) Optical image of a typical 1T-TaS$_2$ flake in the two-terminal configuration. c) Schematic of the charge density wave (CDW) phases of 1T-TaS$_2$ near RT. In the nearly commensurate (NC) phase, Star-of-David supercells (see inset) form hexagonal domains. Above T$_C$, the CDW becomes incommensurate (IC). d) Current-voltage (*IV*) characteristic of Device 1 under zero applied strain. The NC-to-IC phase transition, driven by Joule heating, occurs at the threshold voltage V$_{th}$, appearing as an abrupt current increase in curve I, with hysteresis observed upon reducing the bias in II (similarly for positive voltages in III and IV). e) Strain-tuning of the NC-to-IC transition in Device 2. Tensile strain increases V$_{th}$, while compressive strain decreases V$_{th}$. For visual clarity, only the forward sweeps are shown.

Figure 1a illustrates the experimental concept, where uniaxial strain is applied to a 1T-TaS$_2$ device fabricated on a flexible polycarbonate substrate using a four-point bending setup[55]. A mechanically exfoliated, thin flake is positioned onto pre-patterned source and drain electrodes, forming the conductive channel (Fig 1b). The four-point bending configuration allows for controlled application of tensile ($\varepsilon_T$) and compressive ($\varepsilon_C$) strain directly during electrical measurements.

At room temperature (T < T$_C$), the 1T-TaS$_2$ exhibits a nearly commensurate charge density wave (NC-CDW), characterized by hexagonal domains of CDW supercells arising from a $\sqrt{13} \times \sqrt{13}$ Star-of-David distortion of the Ta atoms (Fig. 1c). In the NC-CDW state, electronic conduction largely takes place along incommensurate regions between the commensurate CDW domains. As the temperature increases above $T_C \approx 350$ K, the CDW order becomes incommensurate with the underlying lattice, and the material enters a metallic incommensurate CDW (IC-CDW) phase. The proximity of this phase transition to RT makes 1T-TaS$_2$ an ideal platform for controlling CDW order under ambient conditions.

Interestingly, this NC-to-IC phase transition can be driven by Joule heating, as shown in Figure 1d. The current-voltage (*IV*) characteristics of the flake exhibit a sharp, abrupt increase in current around ±0.85 V during the forward voltage sweeps (I and III). This current jump signals the transition to the IC-CDW phase, as the dissipated Joule power raises the flake temperature above $T_C$. When the voltage is swept back down (II and IV), the system returns to the NC-CDW phase at around ±0.72 V. The hysteresis observed between forward and backward sweeps reflects the thermal nature of the transition. This behavior is consistent with previous reports on devices fabricated on rigid substrates[35,36,50,56,57]. All measurements presented here are performed at ambient conditions, with the only source of heating being self-heating, i.e. the electrical power dissipation within the device.

We now explore how uniaxial strain modifies this heating-induced phase transition. To do so, we perform *IV* measurements while systematically applying either tensile or compressive strain to the substrate with a four-point bending setup (see Methods). Figure 1e shows representative *IV* curves of the same device (Device 2) for three different strain conditions: zero strain, tensile strain ($\varepsilon_T = 0.3\%$), and compressive strain ($\varepsilon_C = 0.3\%$). While compressive strain $\varepsilon_C$ is usually defined as negative, in this work we will refer to the absolute value of each type of strain, i.e. we define both $\varepsilon_T$ and $\varepsilon_C$ as positive.

In the three cases, an abrupt jump in the current, marking the NC-to-IC phase transition, is clearly visible. Notably, the threshold voltage at which the phase transition occurs, denoted $V_{th}$, is shifted by the applied strain. In the absence of strain $V_{th} \approx 1.01$ V, while tensile strain increases $V_{th}$, delaying the onset of the phase transition. Conversely, compressive strain reduces $V_{th}$, promoting an earlier transition.

To further investigate this tunability, we systematically study how $V_{th}$ evolves as a function of applied strain. Figure 2 shows *IV* characteristics obtained while progressively increasing the applied tensile and compressive strain. We show here two separate devices for tensile (Device 3) and compressive strain (Device 4). For tensile strain (Figs. 2a–b), we observe a clear, monotonic increase in $V_{th}$ as the flake is stretched. A linear fit to the data up to 0.6% strain yields a gauge factor (for the threshold voltage) of $\Delta V_{th}/\Delta \varepsilon = 55$ mV/%. This behavior is reversible and stable over multiple strain cycles (Fig. 2c) that comprise several hours of measurement time, highlighting the robustness of the phenomenon (see Figure S8 for longer cyclic measurements).

Applying compressive strain produces the opposite effect (Fig. 2d), with $V_{th}$ decreasing monotonically as the flake is compressed. The magnitude of the strain sensitivity (Fig. 2e) under compression is similar to the tensile case, with a threshold-voltage gauge factor of $\Delta V_{th}/\Delta \varepsilon = 57$ mV/%. Within the definition used here, this figure of merit remains positive, indicating that $V_{th}$ decreases as the magnitude of compressive strain increases. As for tensile strain, the device shows good reproducibility over multiple strain cycles (Fig. 2f). We note that in Figs. 2e-f, we depict the absolute values of $V_{th}$ extracted from the negative branch of the *IV* characteristic (I < 0), while panel (d) depicts the positive branch.

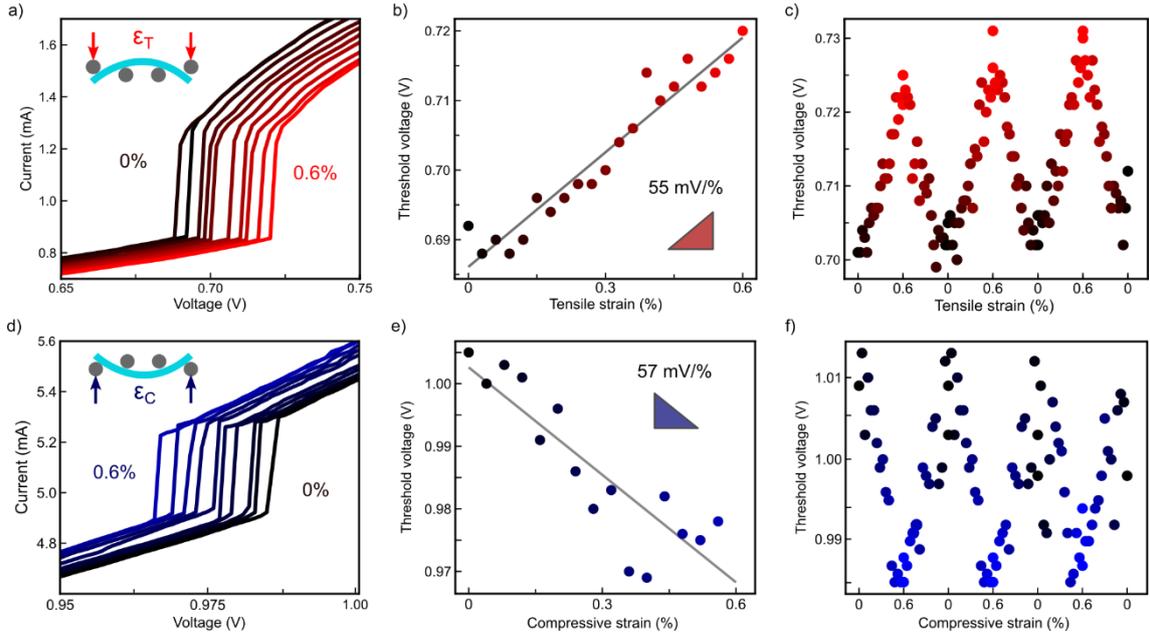

**Figure 2.** a) *IV* characteristics under increasing tensile strain for Device 3. $V_{th}$ rises monotonically with increasing tensile strain. b) Linear fit between $V_{th}$ and $\varepsilon_T$ to extract the threshold voltage gauge factor. c) Cyclic measurement demonstrating stable and reversible control of $V_{th}$ under tensile strain. d) *IV* characteristics under increasing compressive strain for Device 4. $V_{th}$ decreases monotonically with increasing compressive strain. e) Linear fit between $V_{th}$ and $\varepsilon_C$ to extract the threshold voltage gauge factor. $\varepsilon_C$ is defined as negative, keeping the gauge factor positive. f) Cyclic measurement of $V_{th}$ under compressive strain. In panels (e) and (f) the absolute values of $V_{th}$ are extracted from the negative branch of the *IV* curves, in contrast to panel (d).

These results demonstrate that uniaxial strain provides a robust and continuous method to control the NC-to-IC CDW phase transition at RT. We observe this trend consistently across all fabricated devices (additional data in Supp. Info.), where the crystallographic axes of the flakes are randomly aligned with respect to the strain direction. This indicates an in-plane isotopy of the strain modulation of 1T-TaS$_2$, as previously reported[29].

In addition, we explored the effect of biaxial tensile strain on the NC-to-IC phase transition. We fabricated samples in rigid (Si/SiO$_2$) and flexible (PC) substrates and compared the *IV* characteristics as the sample temperature is increased above RT (see Supp. Info.). The thermal expansion of the PC substrate (larger than that of Si/SiO$_2$) exerts biaxial tensile strain on the device[58]. The observed trend in $V_{th}$ agrees with that of uniaxial tensile strain, i.e. $V_{th}$ shifts to larger values as the sample is (biaxially) strained.

Beyond shifting the threshold voltage, uniaxial strain also modifies the flake resistance. We examine the piezoresistance of the NC-CDW phase as it could be linked to the strain tunability of the NC–IC transition. Figures 3a and 3b show the strain-dependent resistance of the NC phase under tensile and compressive strain, respectively: tensile strain increases the resistance approximately linearly, while compression reduces it, yielding a positive gauge factor.

This response likely originates from a geometric effect that typically dominates piezoresistivity in metals. In the NC phase of 1T-TaS$_2$, transport proceeds along a percolative network between commensurate domains[42]. Stretching the flake increases the effective path length and reduces its width (via the Poisson effect), producing the observed positive gauge factor. Other contributions, such as strain-induced changes to the density of states or phonon spectrum, are not expected to significantly affect room-temperature resistance. The piezoresistance of the IC-CDW phase (see Supp. Info.) is consistent in sign and magnitude with the NC-CDW response.

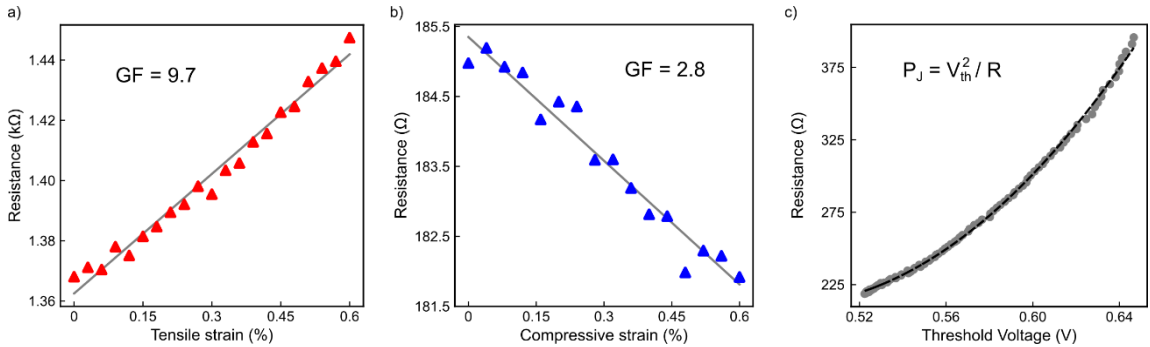

**Figure 3.** a) Resistance of Device 3 in the NC-CDW state under increasing tensile strain. The resistance rises linearly, resulting in a positive gauge factor GF. b) Resistance of Device 4 under increasing compressive strain. The resistance decreases, but the gauge factor remains positive, as compressive strain is defined as negative. c) Quadratic relation between resistance and threshold voltage for Device 5 under uniaxial tensile strain ($R \propto V_{th}^2$), confirming that strain-dependent Joule heating governs the NC-to-IC phase transition.

We quantify the strain sensitivity of the resistance using the piezoresistive gauge factor, defined as GF = $\Delta R/(R_0 \varepsilon)$, where $\Delta R$ is the strain-induced change in resistance, $R_0$ is the resistance under zero strain, and $\varepsilon$ is the strain magnitude. We find GF = 9.7 for tensile strain and GF = 2.8 for compression, in line with typical geometric gauge factors of GF $\approx$ 1-10 for metallic systems[59]. Such piezoresistive effect enables 1T-TaS$_2$ devices to function as bipolar piezoresistive strain gauges, which can detect both tensile and compressive strain continuously. Notably, the devices exhibit the same sign and comparable magnitude of piezoresistance both below and above the NC-to-IC phase transition, enabling operation in either phase. Furthermore, measurements on additional devices reveal larger values for the piezoresistive gauge factor, in the order of GF $\approx$ 100 (see Supp. Info.). Compared to other single-flake 2D piezoresistive strain gauges[60–62], our devices deliver comparable gauge factors, operate at large strain levels and feature far simpler device architectures.

The observed strain dependence of both resistance and $V_{th}$ suggests an intuitive origin for the strain-tunability of the phase transition. The Joule power dissipated in the device, given by $P_J = V^2/R$, depends directly on the flake resistance. Thus, as the applied tensile strain increases $R$, a higher voltage is required to reach the critical power needed to drive the transition to the IC-CDW state. Conversely, compressive strain lowers the resistance, reducing the required threshold voltage. Previous studies have established that the NC–IC transition is primarily driven by Joule heating in the channel[35,56], supporting our hypothesis on the origin for this strain-tunable phase transition.

Assuming that the critical Joule power (i.e. the critical temperature $T_C$) remains constant with strain, this relationship implies $R(\varepsilon) \propto V_{th}^2(\varepsilon)$. To test this hypothesis, we acquire high-resolution *IV* data under incremental strain steps of 0.006% up to 0.6% (see Supp.

Info.) in an additional sample (Device 5). For each strain condition, we extract the corresponding $V_{th}$ and the resistance of the flake in the NC-CDW phase. Plotting these values (Fig. 3c) reveals a clear quadratic relationship between $R(\varepsilon)$ and $V_{th}(\varepsilon)$, confirming that the strain-tunability of the phase transition originates from the piezoresistive modulation of the device resistance.

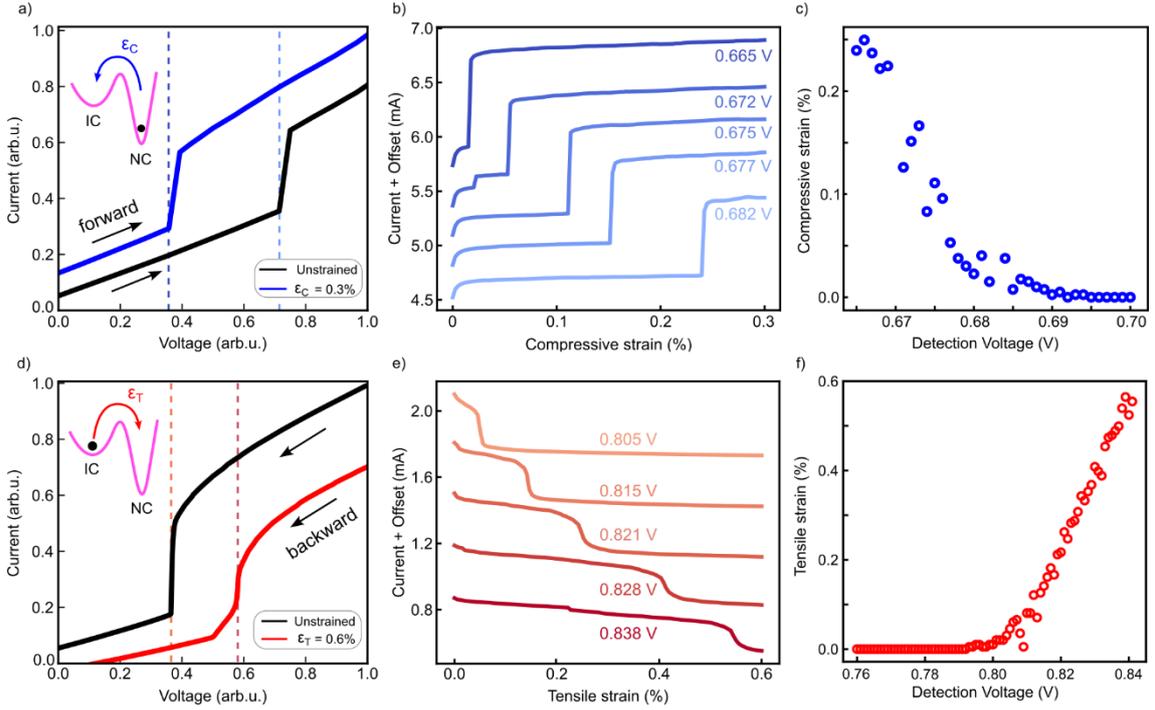

**Figure 4.** a) Operating principle of the CDW-enhanced compressive strain detector. The flake starts in the NC state, biased at a detection voltage $V_{det}$ near the phase transition (within the range indicated by the dashed vertical lines). Compressive strain reduces $V_{th}$, triggering a sharp current increase as the flake switches to the IC state (see inset). b) Abrupt switching of the device upon incremental application of compressive strain, for different detection voltages. c) Programmable detection window for compressive strain, tuned via $V_{det}$. d) Operating principle for the CDW-enhanced tensile strain detector. The flake is pre-biased in the IC phase and held within the threshold voltage window. Tensile strain increases $V_{th}$, prompting the return to the NC state. e) Abrupt switching of the device upon incremental application of tensile strain, for different values of $V_{det}$. f) Programmable detection window for tensile strain detection.

Finally, we leverage the strain-tunable phase transition between CDW phases to demonstrate an alternative mechanism for detection of uniaxial strain at room temperature. We construct threshold-like strain detectors that exploit the destruction (nucleation) of charge order to sense compressive (tensile) uniaxial strain. Exploiting the bi-stability of the *IV* characteristics as an amplification mechanism, similar to snap-through sensors[63], these detectors exhibit exceptional sensitivity and an electrically-tunable detection window.

The detection principle for compressive strain is illustrated in Figure 4a. The device is first initialized in the NC-CDW state under a constant detection bias $V_{det}$, selected close to the strain-dependent threshold voltage $V_{th}$. As discussed earlier, compressive strain lowers $V_{th}$, eventually bringing it below the fixed detection bias. When this condition is met, the dissipated power exceeds the critical value required to drive the NC-to-IC phase transition, resulting in a sharp, measurable increase in current. This threshold-like response is shown in Figure 4b: as compressive strain is gradually increased at constant $V_{det}$, the device undergoes an abrupt switching event when $V_{det} \geq V_{th}$, producing a clear electrical signal. Notably, the strain level at which this transition occurs can be continuously tuned by adjusting $V_{det}$, defining a programmable detection window for compressive strain (dashed lines in Fig. 4a), which in this case extends from 0 to 0.3%.

We quantify the tunability of the detector by plotting the detectable strain as a function of the detection voltage $V_{det}$ (Fig. 4c). A linear fit yields a tunability of $\Delta\varepsilon_C/\Delta V_{det} = 0.0128\%/mV$, allowing precise control over the strain detection threshold. The current jump associated with the phase transition is sizable ($\Delta I \approx 0.75$ mA) and remains largely independent of the specific threshold conditions. Furthermore, the switching occurs abruptly within strain increments of $\Delta\varepsilon_C = 0.0025\%$, resulting in an exceptionally high strain sensitivity of $\Delta I/\Delta \varepsilon_C \approx 300$ mA/%, enabled by the device's intrinsic threshold response.

Detection of tensile strain is also tunable via the choice of the detection bias, as demonstrated in Figures 4e–f. Applying the same definition as before, we extract a detector tunability of $\Delta \varepsilon_T/\Delta V_{det} = 0.0264$%/mV, while the corresponding strain sensitivity is $\Delta I/\Delta \varepsilon_T \approx 112$ mA/%. We note that the IC-to-NC phase transition, which underpins tensile strain detection, is typically less abrupt than the NC-to-IC transition. The reverse transition (IC-to-NC) is governed by slow cooling and thermally-activated nucleation and domain growth[42], resulting in a broader, less sharp switching response. Consequently, the device exhibits reduced sensitivity for tensile strain detection compared to the compressive strain configuration.

As a threshold strain sensor, our CDW-based device achieves a remarkable sensitivity of 300 mA/%, far exceeding other two-dimensional platforms[60–66]. The NC–IC transition produces a sharp change in current, yielding clearly distinguishable signals. Moreover, the strain-detection window can be continuously tuned by adjusting the applied detection voltage, allowing real-time control over the strain range. All in all, our CDW-enhanced strain detector combines high sensitivity with straightforward integration into on-chip device architectures. The main trade-off is that after each detection event, the device remains in the switched state (IC state for compressive strain detection, NC for tensile strain detection). Therefore, the device must be reset before it can detect a new event. Negative feedback circuits[67], which automatically apply a reset signal after each detection event, would be required to support fast, free-running detection of strain fields.

Interpreting uniaxial strain as a change in source–drain separation (L $\simeq$ 25 μm), the device functions as a high-resolution displacement sensor. Under compressive loading, it exhibits a current-displacement sensitivity of ΔI/ΔL ≈ 12 µA/pm (and 4.4 µA/pm under tension), while the displacement-detection voltage slope is $\Delta L/\Delta V_{det}$ ≈ 660 pm/mV (and 320 pm/mV in tension). These parameters enable reliable detection of sub-nanometer motions (down to ΔL = 0.625 nm) within a voltage-controlled sensing window. Compared to leading resistive and piezoresistive displacement sensors, which achieve sensitivities around 1 V/µm[68–70], our CDW detector offers superior sensitivity (600 V/µm assuming a 50 Ω load), albeit over a limited ±15 nm range. Nonetheless, its compact, on-chip footprint and dual operation modes for strain and displacement sensing make it a versatile platform for precision measurements in integrated 2D-material systems.

In conclusion, we have demonstrated reversible, room-temperature control of the NC-to-IC charge density wave phase transition in thin 1T-TaS$_2$ flakes through the application of uniaxial strain. By systematically tuning the strain and monitoring both the transition threshold and device resistance, we experimentally confirm that the strain-dependence of the phase transition originates from the piezoresistive modulation of the flake's resistance, which directly affects the Joule-heating conditions required to drive the transition. This mechanism enables precise and systematic tuning of the NC-to-IC phase transition via tensile and compressive strain, with tunabilities of 55 mV/% and 57 mV/%, respectively.

Leveraging the sharp, threshold-like electrical response at the phase transition, we realize a compact, highly sensitive strain and displacement detector (~ 0.1-0.3 A/% and ~10 µA/pm) with an electrically-programmable detection window. Notably, one can sense sub-nanometer displacements at room-temperature in an on-chip architecture. Compared to conventional approaches, this platform offers a unique combination of sensitivity, tunability, and device simplicity, made possible by the intrinsic properties of CDW order in 1T-TaS$_2$. Beyond strain sensing, these results establish strain-tunable CDW devices as promising candidates for threshold-driven functionalities in emerging technologies such as neuromorphic computing[71,72] and phase-switch electronics[73,74]. Towards practical devices, we note that polymer encapsulation[75] presents a promising avenue to improve further the strain transmission and device stability over time.

MATERIALS AND METHODS

**Device Fabrication**

The 1T-TaS$_2$ crystals were purchased from HQ Graphene (HQ Graphene, The Netherlands). and exfoliated using commercial scotch tape. To achieve thin crystallites suitable for device fabrication, the cleaved crystals were transferred to a polydimethylsiloxane (PDMS) carrier substrate (Gel-Film WF ×4 6.0mil, GelPak) by gently pressing the exfoliation tape against the PDMS surface and peeling it away slowly. This process ensures that a variety of flake thicknesses are available for further selection.

The identification of suitable flakes was carried out using transmission-mode optical microscopy, where semi-transparent flakes were chosen, with thicknesses around ∼ 30 nm. The NC-to-IC phase transition in 1T-TaS$_2$ is known to be largely thickness independent[34,71]. We observe that thicker samples tend to show smaller voltage jumps at the phase transition. As the flake thickness increases, interlayer coupling becomes more important, and we hypothesize that the NC–IC switching dynamics becomes less abrupt as additional conduction channels (including out-of-plane channels) become available.

The selected flakes were then transferred from the PDMS carrier substrate to the final device substrate using an all-dry deterministic transfer method[76,77]. The transfer process was conducted under an optical zoom lens system to ensure proper alignment of the flake over the electrodes. The final substrate consists of a 250 µm thick polycarbonate sheet (Modulor, Article No.0262951) with pre-patterned source and drain electrodes. The polycarbonate substrate, chosen for its flexibility, allowed for the application of controlled strain via mechanical bending. The electrodes were fabricated by evaporating a 45 nm layer of gold onto the polycarbonate substrate, using a 5 nm titanium layer as an adhesion promoter. Both layers were deposited using e-beam evaporation through a commercially available shadow mask (Ossila, Product Code E291) to define the electrode pattern.

**Strain Application and Electrical Measurements**

In a four-point bending geometry, uniaxial strain is induced in the sample by applying a force at the sides of the flexible chip (at the loading pins) while supporting it at the center (supporting pins). Depending on the sign of the applied uniaxial strain, the loading pins apply the force downwards (tensile) or upwards (compressive). The configuration for tensile uniaxial strain is pictured in Figure 1a, while the configuration for compressive strain is shown as an inset in Figure 2d. This configuration creates a bending moment, resulting in a curvature of the sample. For equidistant pins, the strain ($\epsilon$) on the surface of the sample can be expressed as:

$$\epsilon = 27 \cdot t \cdot \Delta / 5d^2,$$

where t is the thickness of the sample, $\Delta$ is the vertical deflection at the midpoint and d is the distance between any two consecutive pins.

Uniaxial strain was applied using a motorized bending setup capable of very precise displacements of the pivotal points of the bending apparatus. The strain levels ranged from 0.0% to 1%, and were previously calibrated by following the protocol described in previous work[55,78]. In the four-point bending used to apply uniaxial tension and compression; reversing the strain sign requires remounting the sample from the tensile to the compressive configuration. During this process, inadvertent strain can induce flake slippage or fracture (see Supp. Info.). To avoid this artifact, we focus on separate devices for tensile and compressive measurements, in order to study their pristine properties. We note that, among other methods to apply strain to transport devices, four-point bending represents a simple, yet powerful method that enables us to dynamically apply strain in a repeatable manner (see Figure S8). In Table 1 of the Supp. Info., we provide a brief comparison with other established strain methods that are compatible with electrical transport.

Electrical measurements were carried out using a Keithley 2450 source meter unit to perform current-voltage (*IV*) sweeps while progressively increasing strain. The CDW transition was induced via Joule heating, with the threshold voltage corresponding to the point at which the current exhibited a sudden increase, signaling the phase transition.

**Strain-Induced Switching**

To investigate strain-induced switching, we fixed the voltage within the transition range, based on the *IV* characteristics, and monitored the source-drain current as a function of strain. This allowed us to observe abrupt current changes associated with strain-induced modifications to the CDW phase. In the case of tensile strain, as the transition moves to higher voltages with increasing strain, the flake is pre-biased to the high-temperature IC-CDW phase and we exploit the hysteretic nature of the phase transition. Thus, we detect switching from the IC-CDW phase to the NC-CDW upon application of tensile strain.

**Biaxial strain measurements**

Biaxial strain measurements were performed in an Attodry800 cryostat under fixed pressure of $10^{-3}$ mbar. *IV* characteristics were measured using a National Instruments NI USB-6343 data acquisition board to apply the source-drain voltage and record the current flowing through the sample, upon amplification by a Femto DLPCA-200 current amplifier. The *IV* sweeps were performed at a rate of approximately 21 mV/s for both samples, measuring one after the other once the target temperature was reached. The fabrication of samples on rigid Si/SiO$_2$ substrates followed the same procedure detailed above.

**Use of AI language models:**

The instrumentation control software for the strain-electrical measurements was developed with AI assistance, following an autonomous-instrumentation workflow[79].

ChatGPT (GPT-4o, OpenAI's large-scale language-generation model) has been used to improve the English grammar and writing style of this manuscript. The authors have reviewed, edited, and revised the ChatGPT generated texts to their own liking and take ultimate responsibility for the content of this publication.

**Content in the Supporting Information file**.

The Supporting Information includes: details on the sample fabrication and the experimental techniques, images of the flake identification and stacking, additional datasets under tensile and compressive uniaxial strain, additional data and discussion on the effects of biaxial strain, discussion on the piezoresistive origin of the transition tunability, examples of multi-step transitions for flakes with inhomogeneous thickness, additional datasets on longer strain cycles, additional data on devices featuring larger gauge factors, additional data on the piezoresistive effect in the IC-CDW phase, additional data and images on the mechanical stress and fracture of the devices. Lastly, we provide additional datasets on the detection of tensile and compressive uniaxial strain, as well as a table comparing other methods for the application of strain that are compatible with transport measurements.

AUTHOR INFORMATION

Corresponding Author

Rafael Luque Merino

rafael.luque@csic.es

Andrés Castellanos-Gómez

andres.castellano@csic.es

Author Contributions

R. L. M., F. C. and E. H-G. were responsible for sample fabrication and conducting the experimental measurements, contributing to the investigation, data collection, and analysis necessary for the study.

M. R. C. and R. F. contributed to the conceptualization of the project and the data curation.

A. C-G. contributed to the conceptualization, funding acquisition, project administration, supervision, methodology development, and the drafting and revision of the manuscript, providing leadership in experimental planning, securing resources, and overseeing all stages of the research process as the thesis supervisor and group leader.

R. L. M. and A. C-G. wrote the manuscript with input from all co-authors. All authors have given approval to the final version of the manuscript.

Funding Sources

A. C-G. acknowledges funding by the Ministry of Science and Innovation (Spain) through the projects PDC2023-145920-I00 and PID2023-151946OB-I00 and the European Research Council through the ERC-2024-PoC StEnSo (grant agreement 101185235) and the ERC-2024-SyG SKIN2DTRONICS. R. L. M. and A. C-G Acknowledge the Severo Ochoa Centres of Excellence program through Grant CEX2024-001445-S. M.R.C. acknowledges funding from PID2023-146354NB-C44, (funded by MICIU/AEI/10.13039/501100011033, and from EU FEDER). MRC and EHG from CNS2023-145151 (funded by MICIU/AEI/10.13039/501100011033 and from EU

NextGenerationEU/PRTR). R. C. acknowledges funding from Ministero dell'Università e della Ricerca (MUR) through the PRIN Grant 2D-PentaSensing (P2022HT3FF).